\documentclass[twocolumn, aps, preprintnumbers, showpacs, prl]{revtex4}
\usepackage{graphicx}
\usepackage{times}
\usepackage{epsfig}
\usepackage{amsmath}
\usepackage{amsfonts}
\usepackage{amssymb}
\usepackage{url}
\usepackage{hyperref}
\usepackage{subfigure}
\newcommand{\beqa}{\begin{eqnarray}} 
\newcommand{\eeqa}{\end{eqnarray}} 
\newcommand{\beq}{\begin{equation}} 
\newcommand{\eeq}{\end{equation}}

\newcommand{\be}{\begin{equation}}
\newcommand{\ee}{\end{equation}}
\newcommand{\bea}{\begin{eqnarray}}
\newcommand{\eea}{\end{eqnarray}}
\newcommand{\bref}[1]{(\ref{#1})}
\begin{document}
\preprint{KEK-TH-1283}
\pagestyle{plain}
\title{Enhanced production of TeV right-handed neutrinos\\
through the large magnetic moment}
\author{Tatsuru Kikuchi}
\email{tatsuru@post.kek.jp}
\affiliation{Theory Division, KEK,
1-1 Oho, Tsukuba, 305-0801, Japan.}
\date{\today}
\begin{abstract}
We examine the effects of magnetic moments of right-handed neutrinos, 
whose masses are set at around TeV scale, then it is plausible to have
a large enhancement for the production cross section of TeV scale
right-handed neutrinos though the Drell-Yan process,
$e^+ e^- \to \gamma, Z^* \to N_i N_j~(i \neq j)$,
which is within the reach of the future linear collider (ILC).
\end{abstract}
\maketitle
\section{Introduction}
The strong evidence of neutrino oscillations from the solar, atmospheric as well as
long baseline accelerator and reactor neutrino measurements implies finite neutrino
masses and mixings, and this is also the evidence of new physics beyond the standard model (SM).

The couplings of neutrinos with the photons are generic consequences of finite neutrino masses, 
and are one of the important intrinsic neutrino properties to explore.
The study of neutrino magnetic moments is, in principle, a way to distinguish between 
Dirac and Majorana neutrinos since the Majorana neutrinos can only have flavor changing,
transition magnetic moments while the Dirac neutrinos can only have flavor conserving one.

However, in the standard electroweak theory extended to include the Dirac-type neutrino masses,
\be
{\cal L}_{{\rm SM} + {\rm Dirac}-N} = {\cal L}_{\rm SM} + y_\nu \overline{N} L H +h.c. 
\ee
that gives the usual Dirac-type neutrino masses as $m_\nu = y_\nu v$ ($v=174$ GeV),
it has been shown in \cite{Fujikawa:1980yx} that 
\be
\mu_\nu = \frac{3 e G_F m_\nu}{8\pi^2 \sqrt{2}} = 3.2 \times 10^{-19} \left(\frac{m_\nu}{\rm meV}\right)\, \mu_B, 
\ee
which is strongly suppressed owing to the chiral symmetry and the GIM cancellations and 
a consequence of small neutrino masses. 

From the experimental side of view, laboratory bounds on the magnetic moments of neutrinos 
are obtained via elastic $\nu-e$ scattering, where the scattered neutrino is not observed.
The most several constraint at the laboratory experiments is given by a reactor experiment \cite{Wong:2006nx}
which gives
\be
\mu_\nu < 0.74 \times 10^{-10} \, \mu_B ~(\mbox{$90$ \% C.L.} )
\ee
In general, the magnetic moments of neutrinos are described by the following dipole operator
\be
{\cal L}_{\rm int} = \mu_\nu \overline{\nu}_L \sigma_{\mu \nu} \nu_R F^{\mu \nu} +h.c. \;,
\label{dipole}
\ee
If the neutrinos are Majorana particles, one can only have a flavor changing dipole operator
\be
{\cal L}_{\rm int} = \mu_\nu^{ij} \nu_L^i C^{-1} \sigma_{\mu \nu} \nu_L^j F^{\mu \nu} +h.c. ~(i \neq j) \;.
\label{dipole}
\ee
The former is the usual Dirac-type magnetic moment, while the latter, which violates lepton number by two units,
is referred to as 'transition magnetic moment'.
Note that both of lepton flavor and CP violations are necessary for the non-zero transition magnetic moment
to be induced (except the CP invariant case with relative CP-phase $\pi$). 
For Majorana neutrinos in the standard model extended to incorporate the see-saw mechanism \cite{yanagida}, 
The transition magnetic moment is found to be \cite{Fukugita:2003en}:
\bea
 \mu_\nu^{ij}=\frac{3 e G_F (m_i+m_j)}{16 \pi^2 \sqrt{2}}
 \sum_{\alpha =e, \mu, \tau}
 {\rm Im} \left[U^{\dagger}_{j \alpha} 
 \left(\frac{m_\alpha}{M_W} \right)^2
 U_{\alpha i}  \right] ,
 \label{SM}
\eea
where $m_i$ is the mass eigenvalue of the i-th generation neutrino, 
$U$ is the MNS lepton mixing matrix, and $M_W$ is the weak boson mass. 
Note that in order to evaluate the transition magnetic moment we need informations of the absolute values 
of neutrino masses and all the elements in $U$ including three CP-phases.

In a very general class of models, one usually finds a relation between the neutrino masses and the magnetic
moments of neutrinos \cite{Babu:1990wv}.
This is because the magnetic moment operator \bref{dipole} has the same chiral structure as
the neutrino mass operator. Thus, if in the diagrams of the neutrino magnetic moment the external 
photon line is removed, it also contributes to the neutrino mass, in general,
\be
\mu_\nu \approx \frac{m_\nu m_e}{\tilde{m}^2}  \mu_B      \;,
\label{scaling}
\ee
where $\tilde{m}$ is the mass of heavy particle that circulate in the loop. In the case of SM, this scale
corresponds to the mass of W, and in the case of supersymmetric extension of the SM (SSM),
it represents a typical mass of SUSY particles' masses \cite{Fukuyama:2003uz}.
In most models which are some extension of the standard model, the magnetic moment of neutrinos is
always proportional to the neutrino mass, and it can not become large enough to be detectable.

On the other hand, in more general models there is no longer a proportionality 
between neutrino mass and its magnetic moment, even though only massive neutrinos
have non-vanishing magnetic moments without fine tuning. 
Indeed, there are some class of models which lead to a large amount of magnetic moment of
neutrinos in \cite{Babu:1989wn, Babu:1989px} where a global symmetry is used and in 
\cite{Voloshin:1987qy, Barbieri:1988fh, Ecker:1989ph, Chang:1990uga, Choudhury:1989pw}
where the gauge symmetry is extended.

\section{Magnetic moment of right-handed neutrinos}
The important notice of this paper is that if the scaling rule of magnetic moment of
neutrinos \bref{scaling} is applied to the heavy right-handed neutrinos,
they can, in general, have a large amount of magnetic moment evading the chiral
suppression.
The basic idea to have a large magnetic moment for the right-handed neutrinos
is motivated by the work in \cite{Sher:2001rk, Sher:2002ij}, where the possibility
to have a large electric dipole moment (EDM) for heavy leptons have been discussed.

In this work, we consider the effects of TeV scale right-handed neutrinos.
It has received some interests on the study of electroweak-scale scale right-handed neutrinos,
see, for instance, Ref. \cite{Hung:2006ap}.
The most generic effective Lagrangian approach to study the effects of heavy Majorana 
neutrinos $N$ with sub-TeV masses has recently been done in Ref. \cite{delAguila:2008ir}.

Now we consider to include the Yukawa coupling of the right-handed neutrinos and
the Majorana mass terms for the right-handed neutrinos
which violates the lepton number by two units, 
\bea
{\cal L}_{{\rm SM} + {\rm Majorana}-N} &=& {\cal L}_{\rm SM} + (y_\nu \overline{N} L H +h.c. )
\nonumber\\
&+& \frac{1}{2} M_i N^i C^{-1} N^i
\eea
that gives the effective left-handed neutrino masses in the form of well-known see-saw formula \cite{yanagida}
as $m_\nu = m_D^T M^{-1} m_D$ with $m_D = y_\nu v$ after integrating out the right-handed neutrinos.
From the see-saw formula, one can estimate the size of the neutrino Yukawa coupling in case of TeV right-handed
neutrino:
\be
y_\nu^2 = 3 \times 10^{-11} \left(\frac{m_\nu}{1~{\rm eV}} \right)  \left(\frac{M}{1~{\rm TeV}} \right)  \;.
\ee
So, it is very tiny but it is remarkable to say that it is just the same order for the electron Yukawa coupling.

If we include the right-handed neutrinos as Majorana particles, 
one can again have a flavor changing dipole operator
\be
{\cal L}_{\rm int} = \mu_N^{ij} N^i C^{-1} \sigma_{\mu \nu} N^j F^{\mu \nu} +h.c. ~(i \neq j) \;.
\label{dipole}
\ee
As similar to the case of left-handed neutrinos, 
the transition magnetic moment for the right-handed neutrinos
is found to be
\bea
\mu_N^{ij} &=& \frac{e y_\nu^2 (M_i+M_j)}{16 \pi^2 m_h^2}
 \sum_{\alpha=e,\mu,\tau} {\rm Im} \left[V^{\dagger}_{j \alpha} \left(\frac{m_\alpha}{m_h} 
 \right)^2 V_{\alpha i} \right] 
\nonumber\\
&\sim &
3 \times 10^{-18} \times \left(\frac{m_\nu}{1~{\rm eV}} \right)  \left(\frac{M}{1~{\rm TeV}} \right)^2 \left(\frac{100~{\rm GeV}}{m_h} \right)^2 \mu_B
\nonumber\\
&&(i\neq j) \;,
\eea
where $m_h$ is the Higgs boson mass, 
$M_i$ is the mass eigenvalue of the i-th generation right-handed neutrino,
and $V$ is a unitary matrix which diagonalize the mass matrix of the right-handed Majorana neutrinos. 

\section{Cross section of the right-handed neutrinos}
A discussion of the differential cross section for a heavy charged lepton can be found in Ref. \cite{Sher:2001rk}. 
Here we are interested in right-handed neutrino production. The differential cross-section for the process,
$e^+ e^- \to \gamma,\,Z^* \to N_i N_j~(i \neq j)$, is given by
\bea
{d\sigma\over d\Omega}&=&
{\alpha^2\over 4s}\sqrt{1-{4M^2\over s}}\left( A_1 + {1\over 8\sin^42\theta_W} P_{ZZ}\ A_2 
\right.
\nonumber\\
&+&\left.
{(1-4\sin^2\theta_W)\tan\theta_W\over \sin^22\theta_W} P_{\gamma Z}\ A_3\right) \;,
\nonumber\\
\eea
where
\bea
A_1 &=& D^2 \, s \, \sin^2\theta \left(1+{4 M^2\over\ s} \right) \;,
\nonumber\\
A_2 &=& 1+\cos^2\theta -{4 M^2\over s}\sin^2\theta + 8 C_V \cos \theta
\nonumber\\
&+& D^2 \, s \, \tan^2 \theta_W \left[\sin^2\theta +{4M^2\over s} \left(1+\cos^2\theta \right) \right] \;,
\nonumber \\
A_3 &=& 4 D^2\, s \, \left[\sin^2\theta+ {4 M^2\over s} \left(1+\cos^2\theta \right) \right] \;,
\nonumber\\
P_{ZZ}&=&{s^2\over (s-M^2_Z)^2+\Gamma^2M^2_Z} \;,
\nonumber\\
P_{\gamma Z} &=& {s(s-M^2_Z)\over (s-M^2_Z)^2+\Gamma^2M^2_Z} \;,
\eea
with $D=\mu_N^{ij}$, $C_V={1\over 2}-2\sin^2\theta_W$, and
we have dropped the numerically negligible $C_V^2$ terms, for simplicity.

\begin{figure}[h]
\begin{center}
\includegraphics[width=.8\linewidth]{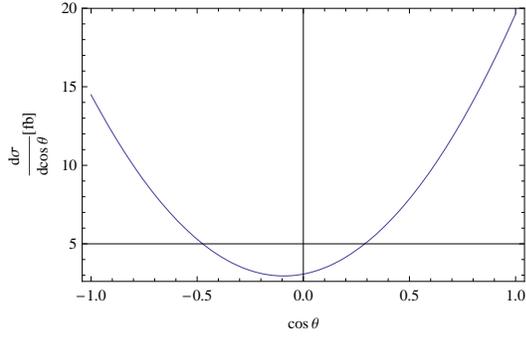}
\end{center}
\caption{
The differential cross section for the process, $e^+ e^- \to \gamma,\,Z^* \to N_i N_j~(i \neq j)$,
is shown as a function of scattering angle $\cos \theta$. 
Here we have fixed $M = 200$ GeV and $\sqrt{s} = 500$ GeV.
}
\label{crosssection0}
\end{figure}

In Fig. \ref{crosssection0}, it is shown the differential cross section for the process, 
$e^+ e^- \to \gamma, Z^* \to N_i N_j~(i \neq j)$, 
for a given heavy Majorana mass scale $M = 200$ GeV and a fixed center of
collider energy $\sqrt{s} = 500$ GeV as a function of scattering angle $\cos \theta$.

\begin{figure}[h]
\begin{center}
\includegraphics[width=.8\linewidth]{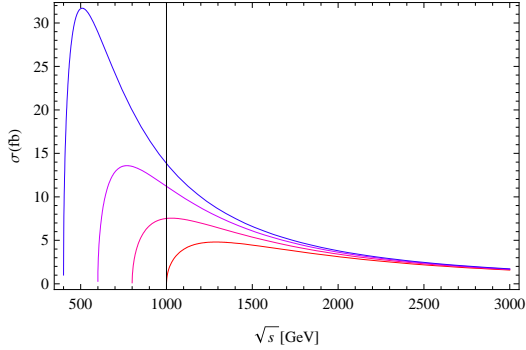}
\end{center}
\caption{
The total cross section for the process, $e^+ e^- \to \gamma,\,Z^* \to N_i N_j~(i \neq j)$,
is shown as a function of center of collider energy $\sqrt{s}$.
Here we have varied $M = 200,~300,~400,~500$ GeV from the top to the bottom curves.
}
\label{crosssection1}
\end{figure}

In Fig. \ref{crosssection1}, it is shown the total cross section for the process,
$e^+ e^- \to \gamma, Z^* \to N_i N_j~(i \neq j)$,
for varied heavy Majorana mass scales $M = 200,~300,~400,~500$ GeV 
as a function of center of collider energy $\sqrt{s}$.

\begin{figure}[h]
\begin{center}
\includegraphics[width=.8\linewidth]{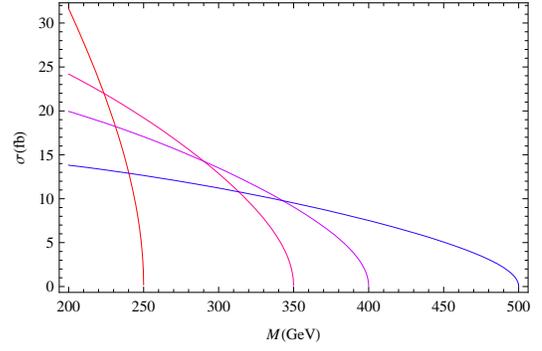}
\end{center}
\caption{
The total cross section for the process, $e^+ e^- \to \gamma,\,Z^* \to N_i N_j~(i \neq j)$,
is shown as a function of heavy Majorana mass scale $M$. 
Here we have chosen $\sqrt{s} = 500, 700, 800, 1000$ GeV from the left to the right.
}
\label{crosssection2}
\end{figure}

In Fig. \ref{crosssection2}, it is shown the total cross section for the process,
$e^+ e^- \to \gamma, Z^* \to N_i N_j~(i \neq j)$,
for varied center of collider energies $\sqrt{s} = 500,~700,~800,~1000$ GeV 
as a function of heavy Majorana mass scale $M$.

In these plots, we have used an approximation that the final state right-handed neutrinos 
have almost the same masses with each other, which is denoted by $M$. 
It can be seen that the total cross section for the production of TeV right-handed neutrinos
can reach a few fb, $\sigma \sim 5$ fb.


After the production of a right-handed neutrino, it decays into a left-handed neutrino ($\nu_i$) 
and a Higgs boson ($h$), $N_i \to \nu_i + h$. So, in total, we have two different flavor neutrinos
and two Higgs bosons although the Higgs would subsequently decay into the SM particles 
producing some jets. Therefore, finally we would have some jets ($jj$) plus a missing energy
originated from a pair of neutrinos having different flavors.

\newpage

\section{Conclusion}
We have considered the magnetic moments of right-handed neutrinos, 
whose masses are set at around TeV scale.
Because of the scaling rule of magnetic moment of neutrinos, the heavy 
right-handed neutrinos can, in general, have a large amount of magnetic moments
evading a chiral suppression.
Such large magnetic moments can enhance the production cross section of TeV scale
right-handed neutrinos though the Drell-Yan process,
$e^+ e^- \to \gamma, Z^* \to N_i N_j~(i \neq j)$, 
which is within the reach of the future linear collider (ILC).

\vspace{1cm}
\acknowledgments
The work of T.K. is supported by the Research Fellowship of the Japan Society 
for the Promotion of Science (\#1911329).
We would like to thank M. Nagai for stimulating discussions.

\newpage


\begin{thebibliography}{9}
\bibitem{Fujikawa:1980yx}
  K.~Fujikawa and R.~Shrock,
  Phys.\ Rev.\ Lett.\  {\bf 45}, 963 (1980).
\bibitem{yanagida}
P.~Minkowski,
Phys.\ Lett.\  B {\bf 67}, 421 (1977);
%
T.~Yanagida, in {\it Proceedings of the Workshop on the Unified Theories and Baryon
Number in the Universe,} Tsukuba, Japan, Feb. 13-14, 1979, p.95, 
(eds. O. Sawada and A. Sugamoto, KEK Report No. 79-18, Tsukuba); 
Prog.\ Theor.\ Phys.\  {\bf 64}, 1103 (1980);
%
P.~Ramond, CALT-68-709, Feb 1979. 21pp.
Invited talk given at Sanibel Symposium, Palm Coast, Fla., Feb 25-Mar 2, 1979.
Published in *Paris 2004, Seesaw 25* 265-280
hep-ph/9809459;
%
S.~Glashow, in {\it Proceedings of the Carg\'ese Summer Institute on Quarks and Leptons,}
Carg\'ese, July 9-29, 1979, eds. M. L\'evy et. al, (Plenum, 1980, New York), p.707; 
%
R.~N.~ Mohapatra and G.~Senjanovi\'c,
Phys.\ Rev.\ Lett.\ {\bf 44}, 912 (1980).
\bibitem{Fukugita:2003en}
  M.~Fukugita and T.~Yanagida,
  {\it ``Physics of neutrinos and applications to astrophysics,''}
{\it  Berlin, Germany: Springer (2003) 593 p}
\bibitem{Wong:2006nx}
  H.~T.~Wong {\it et al.}  [TEXONO Collaboration],
  Phys.\ Rev.\  D {\bf 75}, 012001 (2007)
  [arXiv:hep-ex/0605006].
\bibitem{Babu:1990wv}
  K.~S.~Babu and R.~N.~Mohapatra,
  Phys.\ Rev.\  D {\bf 42}, 3778 (1990).
\bibitem{Fukuyama:2003uz}
  T.~Fukuyama, T.~Kikuchi and N.~Okada,
  Int.\ J.\ Mod.\ Phys.\  A {\bf 19}, 4825 (2004)
  [arXiv:hep-ph/0306025].
\bibitem{Babu:1989wn}
  K.~S.~Babu and R.~N.~Mohapatra,
  Phys.\ Rev.\ Lett.\  {\bf 63}, 228 (1989).
\bibitem{Babu:1989px}
  K.~S.~Babu and R.~N.~Mohapatra,
  Phys.\ Rev.\ Lett.\  {\bf 64}, 1705 (1990).
\bibitem{Voloshin:1987qy}
  M.~B.~Voloshin,
  Sov.\ J.\ Nucl.\ Phys.\  {\bf 48}, 512 (1988)
  [Yad.\ Fiz.\  {\bf 48}, 804 (1988)].
\bibitem{Barbieri:1988fh}
  R.~Barbieri and R.~N.~Mohapatra,
  Phys.\ Lett.\  B {\bf 218}, 225 (1989).
\bibitem{Ecker:1989ph}
  G.~Ecker, W.~Grimus and H.~Neufeld,
  Phys.\ Lett.\  B {\bf 232}, 217 (1989).
\bibitem{Chang:1990uga}
  D.~Chang, W.~Y.~Keung and G.~Senjanovic,
  Phys.\ Rev.\  D {\bf 42}, 1599 (1990).
\bibitem{Choudhury:1989pw}
  D.~Choudhury and U.~Sarkar,
  Phys.\ Lett.\  B {\bf 235}, 113 (1990).
\bibitem{Sher:2001rk}
  M.~Sher,
  Phys.\ Rev.\ Lett.\  {\bf 87}, 161801 (2001)
  [arXiv:hep-ph/0105340].
\bibitem{Sher:2002ij}
  M.~Sher and S.~Nie,
  Phys.\ Rev.\  D {\bf 65}, 093018 (2002)
  [arXiv:hep-ph/0201220].
\bibitem{Hung:2006ap}
P.~Q.~Hung,
  Phys.\ Lett.\  B {\bf 649}, 275 (2007)
  [arXiv:hep-ph/0612004];
  Frascati Phys.\ Ser.\  {\bf 44}, 313 (2007)
  [arXiv:0706.2753 [hep-ph]];
  Phys.\ Lett.\  B {\bf 659}, 585 (2008)
  [arXiv:0711.0733 [hep-ph]];
  Nucl.\ Phys.\  B {\bf 805}, 326 (2008)
  [arXiv:0805.3486 [hep-ph]];
 A.~Aranda, J.~Hernandez-Sanchez and P.~Q.~Hung,
  arXiv:0809.2791 [hep-ph].
\bibitem{delAguila:2008ir}
  F.~del Aguila, S.~Bar-Shalom, A.~Soni and J.~Wudka,
  arXiv:0806.0876 [hep-ph].
%
\end{thebibliography}
\end{document}